\documentclass[preprint,aps]{revtex4}

\usepackage{graphicx}
\usepackage{dcolumn}
\usepackage{bm}
  
\begin{document}
\title{Electronic Structure of Calcium Hexaboride within 
the Weighted Density Approximation}
\author{Zhigang Wu$^1$, D.J. Singh$^2$ and R.E. Cohen$^1$}
\affiliation{$^1$Geophysical Laboratory, Carnegie Institution of Washington,
5251 Broad Branch Road, NW, Washington, DC 20015 \\
$^2$Center for Computational Materials Science,
Naval Research Laboratory, Washington, DC 20375}

\date{\today}

\begin{abstract}
We report calculations of the electronic structure of CaB$_6$
using the weighted density approximation (WDA) to density functional theory.
We find a semiconducting band structure with a sizable gap, in contrast
to local density approximation (LDA)
results, but in accord with recent experimental data.
In particular, we find an $X$-point band gap of 0.8 eV.
The WDA correction of the LDA error in describing the electronic
structure of CaB$_6$ is discussed in terms of the orbital character
of the bands and the better cancelation of self-interactions within the
WDA.
\end{abstract}


\maketitle

The discovery of ferromagnetism in the lightly doped 
CaB$_6$ and SrB$_6$ resulted in considerable experimental and theoretical
work aimed at understanding this unexpected phenomenon.
The reported ferromagnetism is particularly remarkable considering that
the material, Ca$_{1-x}$La$_x$B$_6$, contains no elements with
partially filled $d$ or $f$ orbitals, and even more strangely,
the ferromagnetism is associated with very low doping levels
($<1\%$), and therefore low carrier concentrations, has a very
low moment ($\sim$ 0.07 $\mu_{\rm B}$/La for $x$=0.005),
but a very high Curie temperature, $T_C$ as high as 600K
(Ref. \onlinecite{young99}).
Further, a semiconducting
material, with ferromagnetism of this type, could potentially
be of importance in producing room temperature spintronic devices.
\cite{wolf,tromp2001}

Two different classes of theories were advanced early on in an effort to
explain the strange ferromagnetism. The first was based on
the doping of an excitonic or like system, {\em i.e.}
an unconventional state in the parent, CaB$_6$,
\cite{zhit99,bal2000,bar2000,mura2002}
and the second was based on the ferromagnetic phase of the
dilute electron gas. \cite{young99,cep99,ortiz99}
In the first class, the ferromagnetism arises from La induced doping of
carriers into an excitonic ground state.
This depends on a bare band
structure that is semimetallic or very close to it, so that the
prerequisite instability of CaB$_6$ to an
excitonic state can exist. The second class mentioned
does not depend on such an electronic structure, but rather involves
a ferromagnetic instability of a low density electron gas, though
this class of theories suffers from difficulties obtaining the very
high reported values of $T_C$. More pedestrian explanations for
the observed ferromagnetism involving defects and their interactions
and also extrinsic effects have been advanced more recently.
\cite{gavilano,matsubayashi,matsubayashi2,young2}

This naturally resulted in a strong interest in the electronic structure
of the parent compound CaB$_6$.
First-principles calculations 
within the local density approximation (LDA) predicted
that CaB$_6$ and SrB$_6$ are semimetals
with a small overlap at X point of the simple cubic
Brillouin zone, or very small gap insulators, \cite{mass97,rod2000}
consistent with the excitonic theories being formulated at that time.
However, it is well known that 
the LDA often underestimates band gaps, and in some extreme cases it can
predict a semimetallic band structure for a semiconductor,
one example being Ge.
\cite{fil}
In simple semiconductors, the LDA underestimates of the band gaps are
understood as being associated with an extreme non-analytic
and non-local behavior of density functional theory as
the particle number is changed to include a single conduction
electron, \cite{godby,hybertsen}
implying the need for a self-energy correction, or at least
an orbital dependent potential to obtain a realistic description
of band gaps.

The many-body 
quasiparticle $GW$ approximation \cite{hedin}
is very succesful in describing
the band structures
of insulators.
In particular the self-energy correction
implicit in it corrects most of the band gap errors in simple
semiconductors and other materials as well.
\cite{aryasetiawan}
However, its results for CaB$_6$ have differed
depending on the formulation of $GW$ being used.
Pseudopotential based $GW$ calculations
\cite{tromp2001} yielded a sizable 0.8 eV band gap
at X point, while recent all-electron $GW$ results \cite{kino2002} showed an 
unexpected increase in band
overlap relative to the LDA, perhaps related to the
treatment of deep-core states.
\cite{kotani2001,ku2002}

The electronic structure of CaB$_6$ has also been the subject of some
debate from an experimental point of view, although a semiconducting
electronic structure has become more generally accepted now.
de Haas-van Alphen (dHvA) \cite{good98,hall2001} and Shubnikov de Hass 
(SdH) \cite{aron99} experiments were consistent with a
semimetallic nature, while recent 
angle-resolved photoemission spectroscopy (ARPES) and resonant inelastic 
x-ray scattering (RIXS) measurements
imply that CaB$_6$ is a semiconductor with
a 1.0 (Ref. \onlinecite{souma2003}) - 1.15 
(Ref. \onlinecite{denlinger2002}) eV
band gap.
Recently, Rhyee {\it et\ al}. reported transport, optical, 
and tunneling measurements all consistent with
a 1.0 eV band gap for pure CaB$_6$. They
argued that the previously measured semimetallic
characteristics in CaB$_6$, SrB$_6$, and EuB$_6$ actually originate from boron 
vacancies. \cite{rhyee2003}

We note that, as pointed out by Rodriguez and co-workers,
\cite{rod2000,pic03}
the valence and conduction bands of CaB$_6$ and the related
hexaborides are of very different character, namely B $p$ valence
bands and Ca derived conduction bands. This is reminiscent of
the situation in the rare-earth trihydrides, {\em e.g.} YH$_3$.
Those materials feature a semiconducting ground state, which is
however predicted to be metallic in the LDA due to a band
overlap between nominally occupied H derived states and nominally
unoccupied metal states. In YH$_3$, this error is understood to
be due to incomplete cancelation of self-interaction errors
on the H sites in the LDA. These self-interactions lead to a H band
position that is too high relative to the conduction bands,
which is at least partly corrected in $GW$ calculations.
\cite{eder,vg2,miyake}
It should be noted though that such self-interactions are a local
effect, more akin in some ways to the on-site Coulomb effects that
lead to Mott insulating states in transition metal oxides, than to
the non-local, non-analytic discontinuities in density functional
theory that are important in simple semiconductors. \cite{cohen}
As such, the rare earth trihydrides and perhaps CaB$_6$ are interesting
test cases for theories of correlated electron systems, including
density functional based approaches.

Here,
we present results of first-principles calculations based on the density 
functional theory (DFT) within a non-local density approximation, namely the 
weighted density approximation (WDA). \cite{gunna76,alonso78,gunna79,singh93}
The WDA uses a model pair-distribution function to
incorporate non-local information about the charge density into
the exchange correlation energy and potential, via an exact
expression for the total energy.
Unlike $GW$ and like approaches, the WDA is oriented towards
total energies rather than excitation spectra. As such, not
surprisingly, it improves ground state properties, like lattice
parameters, relative to the LDA.
Like the LDA, the WDA is exact for the total energy of the uniform
electron gas, but unlike the LDA it is also exact for single electron
systems, where it is self-interaction free.
We emphasize that the WDA is a smooth non-pathalogical approximation
to density functional theory, with only short range non-locality.
The WDA exchange correlation potential is a standard orbital
independent (for valence states) potential, which different
orbitals feel differently only through differences in the spatial
distrubutions of their charge densities.
It does not include the discontinuities that are
present in the exact density functional theory and are responsible
for most of the band gap correction in simple semiconductors like Si.
Detailed calculations for YH$_3$ and LaH$_3$ have shown that the WDA
can yield insulating gaps of similar magnitude to the $GW$ approximation
in these hydrides due to the better cancelation of self-interaction
errors, while at the same time improving the quality of
ground state properties. \cite{wu2004}

In the WDA the exchange-correlation (xc) energy 
expression is very similar to the general DFT form, 
\begin{equation}
E^{\rm WDA}_{\rm xc}[n] 
= \int\int \frac{n({\bf r})n({\bf r}^{\prime})}{|{\bf r-r}^{\prime}|}
G[|{\bf r-r}^{\prime}|, \bar{n}({\bf r})] d{\bf r} d{\bf r}^{\prime},
\label{eq1}
\end{equation}
where $G$ is the approximate model
coupling constant averaged pair-distribution function, and $\bar{n}({\bf r})$ 
is the weighted density which can be determined from the sum rule
which states that the charge density in the exchange correlation hole is
unity.
\begin{equation}
\int n({\bf r}^{\prime}) G[|{\bf r-r}^{\prime}|, \bar{n}({\bf r})]
d{\bf r}^{\prime} = -1.
\label{eq2}
\end{equation}
The WDA is exact for both a one-electron system and the uniform electron gas,
independent of the model $G$ used, provided that it is
monotonic and properly normalized.
Because the exact $G$ function is not known,
different forms
of $G$ have been tried, \cite{mazin} from
which we chose the following one
\begin{equation}
G(r,n) = c\{1-{\rm exp}(-[\frac{r}{\lambda}]^{-4})\},
\label{eq3}
\end{equation}
This choice yields good energetics (structural properties) and
band structures for the rare earth trihydrides.
\cite{wu2004} This type of $G$ was derived from the Gunnarsson-Jones  
\cite{gunna80} ansatz, but it has a shorter range than the original 
one. Details of the WDA formalism used in this paper can be found in Refs. 
\onlinecite{wu2004} and \onlinecite{wu2003}.

The calculations were done using a planewave basis with pseudopotentials.
The {\bf k}-grids for the Brillouin zone
samplings during iteration to self-consistency were
as dense as $12 \times 12 \times 12$,
and the energy cutoff
was 121 Ry. \cite{dodpw}
The Ca semicore states were treated as valence states for the
purpose of constructing Troullier-Martins pseudopotentials.
Shell partitioning was employed as discussed in Ref. \onlinecite{singh93}.
The crystallographic structure of CaB$_6$ is simple cubic,
with Ca atoms 
located at cubic corners. The
B atoms form octahedral cages around the cube center. The internal 
parameter $x$ determines the distance
between a B atom and the cubic center. When 
$x=0.207$, the intra- and inter- octahedral B-B distances are equal. Our 
calculated lattice constant $a$ and
internal parameter $x$ within the LDA and WDA are 
shown in Table \ref{tab1}. The WDA results are in excellent agreement with 
experimental data,
while the LDA predicts a lattice constant 1.6\% smaller, which 
is a typical LDA tendency.
Thus, reflecting its origins in density functional theory, where
energy and density are primary variables, the WDA improves upon the
LDA structure.

\begin{figure}
\includegraphics[width=0.78\textwidth]{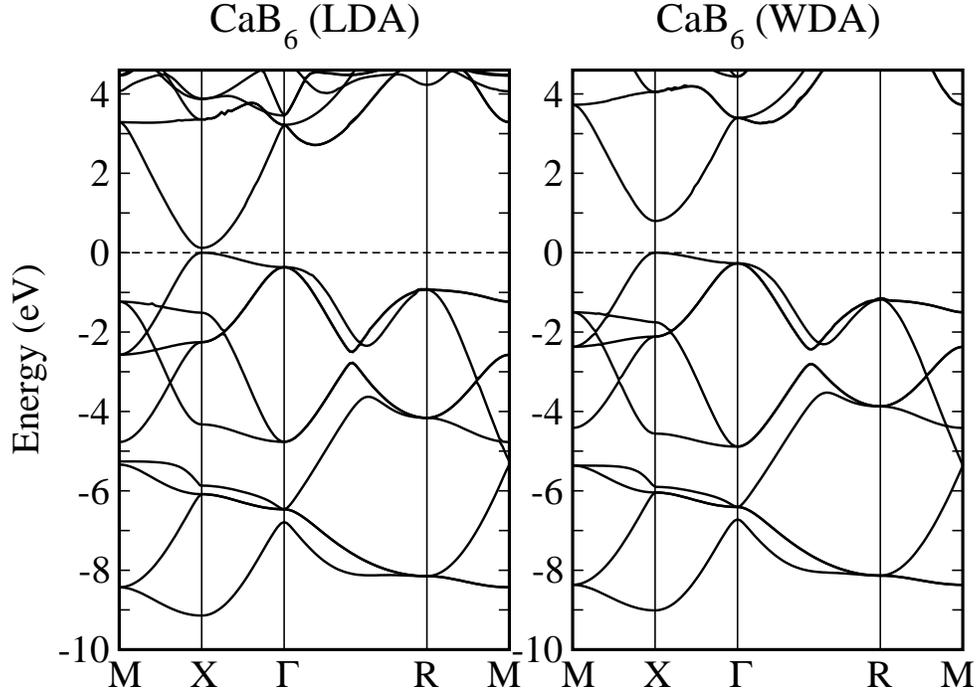}
\caption{\label{fig1} Electronic band structure of CaB$_6$ within
the LDA and WDA.}
\end{figure}

To facilitate comparison of the LDA and WDA
electronic band structures, we used the experimental values of 
$a$ and $x$ for both the LDA and WDA. Fig. \ref{fig1} shows our results
for the band structures of CaB$_6$. The LDA (left panel) predicts a tiny 
band gap (slightly more than 0.1 eV) at the
$X$-point, which agrees with Ref. \cite{kino2002},
while other 
LDA results show a tiny overlap \cite{mass97,rod2000,tromp2001}. This small 
difference in band structure is due to different choices of the LDA scheme,
chemistry (Sr {\em vs.} Ca) and lattice structure as well as
the approximations in using pseudopotentials.
To check this we did well converged general potential linearized
augmented planewave (LAPW) calculations within the LDA
for two structures. The first is
the same experimental structure as used for Fig. \ref{fig1},
and the second is a structure where we relaxed the internal coordinate
$x$ in the LDA keeping the lattice parameter fixed at the experimental
value. This
yielded $x$=0.2025.
With the experimental value of $x$=0.207 the LAPW calculations yielded
a small band gap of 0.2 eV, while at $x$=0.2025 a semimetal with 
a 0.1 eV band overlap is predicted. This implies that the structure
is the key ingredient in reconciling the various results and also
gives an error estimate of $\sim$ 0.1 eV for our pseudopotential
calculations.
On the other hand, the right panel shows that within the WDA, the bands
are shifted,
quite similar to the cases of YH$_3$ and LaH$_3$,
resulting in a sizable band gap of 0.8 eV at X point.
Parallel WDA calculations using the uniform electron gas $G$ function
also yield a gap opening, though by a smaller amount. The band
gap in that case is 0.3 eV.

Thus the WDA, with an appropriate choice of $G$,
predicts CaB$_6$ to be a semiconductor with a gap
comparable to that found in standard $GW$ calculations \cite{tromp2001}
and in experimental
spectroscopic measurements. \cite{souma2003,denlinger2002,rhyee2003}
This shows that there exists a physical approximation to density
functional theory that is continuous and has only short range
non-locality and that gives both a reasonable band gap and
ground state properties (crystal structure) in agreement with
experiment for CaB$_6$.
We expect that WDA calculations
would yield band gap opening in SrB$_6$ as
well.

In summary, WDA calculations
predict that CaB$_6$ is a semiconductor, and the 
calculated band gap is in excellent agreement with recent experiments and 
quasi-particle $GW$ calculations. Thus the WDA improves both
the structural properties and electronic structure of these hexaborides,
implying that the main error in the LDA description is associated with
local self-interaction effects on B rather than the neglect of
the non-analytic
discontinuity in density functional theory across the band gap.

We are grateful for discussions with M.H. Cohen, M. Gupta, R. Gupta,
I.I. Mazin and W.E. Pickett.
DJS is grateful for the hospitality of the University of Paris-Sud
where some of the ideas leading to this work were developed.
Work at the Geophysical Laboratory
was supported by Office of Naval Research (ONR) grants
N000149710052, N00014-02-1-0506 and N0001403WX20028.
Work at the Naval Research Laboratory is supported by ONR.
Calculations were done on the 
Center for Piezoelectrics by Design (CPD) computer facilities.

\pagebreak

\begin{table}
\caption{Theoretical and experimental lattice constant $a$ (\AA) and 
internal parameter $x$ for CaB$_6$.
\label{tab1}}
\begin{ruledtabular}
\begin{tabular}{lccc}
                   &  LDA  &  WDA  &  Experiment \\
\hline
$a$                & 4.078 & 4.131 &  4.146      \\
$x$                & 0.201 & 0.206 &  0.207      \\
\end{tabular}
\end{ruledtabular}
\end{table}



\begin{references}
\bibitem{dodpw}
The calculations were done with a modification of the DoD-PW code.
\end{references}
\end{document}